\documentclass[final,authoryear,3p,times,twocolumn]{elsarticle}

\usepackage{amssymb}

\journal{Planetary and Space Science}

\begin{document}

\begin{frontmatter}


\title{Fast Magnetic Reconnection and Energetic Particle Acceleration}
\author{A. Lazarian$^1$, G. Kowal$^{1,2,3}$, E. Vishniac$^4$ \& E. de Gouveia Dal Pino$^2$}
\cortext[c]{Corresponding author: A. Lazarian, e-mail:lazarian@astro.wisc.edu}
\address[]{\scriptsize
       $^1$Department of Astronomy, University of Wisconsin-Madison, 475 N. Charter St., Madison, WI, 53706, USA,\\
       $^2$Department of Astronomy of IAG, University of S\~{a}o Paulo, Rua do Mat\~{a}o, 1226, S\~{a}o Paulo, SP, 05508, Brazil,\\
       $^3$Astronomical Observatory, Jagiellonian University, Orla 171, 30-244 Krak\'ow, Poland\\
       $^4$Department of Physics and Astronomy, McMaster University, 1280 Main St. W, Hamilton, ON, L8S 4M1, Canada}

\begin{abstract}
Our numerical simulations show that the reconnection of magnetic field becomes
fast in the presence of weak turbulence in the way consistent with the Lazarian
\& Vishniac (1999) model of fast reconnection.  We trace particles within our
numerical simulations and show that the particles can be efficiently accelerated via the first order Fermi acceleration.  We discuss the acceleration arising
from reconnection as a possible origin of the anomalous cosmic rays measured by
Voyagers.
\end{abstract}


\end{frontmatter}


\section{Magnetic Reconnection in Collisionless and Collisional Fluids}

A magnetic field embedded in a perfectly conducting fluid preserves its topology for all time (Parker 79).  Although ionized astrophysical objects, like stars and galactic disks, are almost perfectly conducting, they show indications of changes in topology, ``magnetic reconnection'', on dynamical time scales (Parker 1970, Lovelace 1976, Priest \& Forbes 2002).  Reconnection can be observed directly in the solar corona ( Innes et al 1997, Yokoyama \& Shibata 1995, Masuda et al. 1994), but can also be inferred from the existence of large scale dynamo activity inside stellar interiors (Parker 1993, Ossendrijver 2003).  Solar flares (Sturrock 1966) and $\gamma$-ray busts (Fox et al. 2005, Galama et al. 1998) are usually associated with magnetic reconnection.  Previous work has concentrated on showing how reconnection can be rapid in plasmas with very small collisional rates (Shay et al. 1998, Drake 2001, Drake et al. 2006, Daughton et al. 2006), which substantially constrains astrophysical applications of the corresponding reconnection models.

We note that if magnetic reconnection is slow in some astrophysical environments, this automatically means that the results of present day numerical simulations in which the reconnection is enevitably fast due to numerical diffusivity do not correctly represent magnetic field dynamics in these environments. If, for instance, the reconnection were slow in collisional media this would entail the conclusion that the entire crop of interstellar, protostellar and stellar MHD calculations would be astrophysically irrelevant. 

Here we present numerical evidence, based on three dimensional simulations, that reconnection in a turbulent fluid occurs at a speed comparable to the rms velocity of the turbulence, regardless of either the value of the resistivity or degree of collisionality.  In particular, this is true for turbulent pressures much weaker than the magnetic field pressure so that the magnetic field lines are only slightly bent by the turbulence.  These results are consistent with the proposal by Lazarian \& Vishniac (1999, henceforth LV99) that reconnection is controlled by the stochastic diffusion of magnetic field lines, which produces a broad outflow of plasma from the reconnection zone.  This work implies that reconnection in a turbulent fluid typically takes place in approximately a single eddy turnover time, with broad implications for dynamo activity (Parker 1970, 1993, Stix 2000) and particle acceleration throughout the universe (de Gouveia dal Pino \& Lazarian 2003, 2005, Lazarian 2005, Drake et al. 2006).

Astrophysical plasmas are often highly ionized and highly magnetized (Parker 1970).  The evolution of the magnetic field in a highly conducting fluid can be described by a simple version of the induction equation
\begin{equation}
\frac{\partial \vec{B}}{\partial t} = \nabla \times \left( \vec{v} \times \vec{B} - \eta \nabla \times \vec{B} \right) ,
\end{equation}
where $\vec{B}$ is the magnetic field, $\vec{v}$ is the velocity field, and $\eta$ is the resistivity coefficient.  Under most circumstances this is adequate for discussing the evolution of magnetic field in an astrophysical plasma.  When the dissipative term on the right hand side is small, as is implied by simple dimensional estimates, the magnetic flux through any fluid element is constant in time and the field topology is an invariant of motion.   On the other hand, reconnection is observed in the solar corona and chromosphere (Innes et al. 1997, Yokoyama \& Shibata 1995, Masuda et al. 1994, Ciaravella \& Raymond 2008), its presence is required to explain dynamo action in stars and galactic disks (Parker 1970, 1993), and the violent relaxation of magnetic fields following a change in topology is a prime candidate for the acceleration of high energy particles (de Gouveia Dal Pino \& Lazarian 2003, henceforth GL03, 2005, Lazarian 2005, Drake et al. 2006, Lazarian \& Opher 2009, Drake et al. 2010) in the universe.   Quantitative general estimates for the speed of reconnection start with two adjacent volumes with different large scale magnetic field directions (Sweet 1958, Parker 1957).

The speed of reconnection, i.e. the speed at which inflowing magnetic field is annihilated by ohmic dissipation, is roughly $\eta/\Delta$, where $\Delta$ is the width of the transition zone (see Figure 1).  Since the entrained plasma follows the local field lines, and exits through the edges of the current sheet at roughly the Alfven speed, $V_A$, the resulting reconnection speed is a tiny fraction of the Alfven speed, $V_A\equiv B/(4\pi \rho)^{1/2}$ where $L$ is the length of the current sheet.  When the current sheet is long and the reconnection speed is slow this is referred to as Sweet-Parker reconnection.  Observations require a speed close to $V_A$, so this expression implies that $L\sim \Delta$, i.e. that the magnetic field lines reconnect in an ``X point''.

The first model with a stable X point was proposed by Petschek (1964).  In this case the reconnection speed may have little or no dependence on the resistivity.  The X point configuration is known to be unstable to collapse into a sheet in the MHD regime (see Biskamp 1996), but in a collisionless plasma it can be maintained through coupling to a dispersive plasma mode (Sturrock 1966).  This leads to fast reconnection, but with important limitations.  This process has a limited astrophysical applicability as it cannot be important for most phases of the interstellar medium (see Draine \& Lazarian 1998 for a list of the idealized phases), not to speak about dense plasmas, such as stellar interiors and the denser parts of accretion disks.  In addition, it can only work if the magnetic fields are not wound around each other, producing a saddle shaped current sheet.  In that case the energy required to open up an X point is prohibitive. The saddle current sheet is generic for not parallel flux tubes trying to pass through each other. If such a passage is seriously constrained,
the magnetized highly conducting astrophysical fluids should behave more like Jello rather than normal fluids.  

Finally, the traditional reconnection setup does not include ubiquitous astrophysical turbulence\footnote{The set ups where instabilities play important role include Simizu et al. (2009a,b). For sufficiently large resolution of simulations those set-ups are expected to demonstrate turbulence. Turbulence initiation is also expected in the presence of plasmoid ejection (Shibata \& Tanuma 2001). Numerical viscosity constrains our ability to sustain turbulence via reconnection, however.} (see Armstrong, Rickett \& Spangler 1994, Elmegreen \& Scalo 2004, McKee \& Ostriker 2007, Haverkorn, Lazarian 2009, Chepurnov \& Lazarian 2010). Fortunately, this approach provides another way of accelerating reconnection. Indeed, an alternative approach is to consider ways to decouple the width of the plasma outflow region from $\Delta$.  The plasma is constrained to move along magnetic field lines, but not necessarily in the direction of the mean magnetic field.  In a turbulent medium the two are decoupled, and fluid elements that have some small initial separation will be separated by a large eddy scale or more after moving the length of the current sheet.  As long as this separation is larger than the width of the current sheet, the result will not depend on $\eta$. 

\begin{figure}[!t]
 \begin{center}
\includegraphics[width=1.0 \columnwidth]{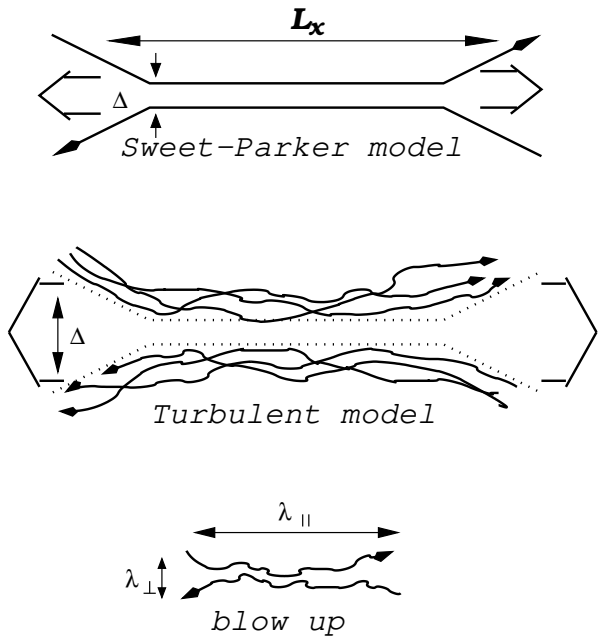}
\caption{{\it Upper plot}:
Sweet-Parker model of reconnection. The outflow
is limited by a thin slot $\Delta$, which is determined by Ohmic
diffusivity. The other scale is an astrophysical scale $L\gg \Delta$.
{\it Middle plot}: Reconnection of weakly stochastic magnetic field according to
LV99. The model that accounts for the stochasticity
of magnetic field lines. The outflow is limited by the diffusion of
magnetic field lines, which depends on field line stochasticity.
{\it Low plot}: An individual small scale reconnection region. The
reconnection over small patches of magnetic field determines the local
reconnection rate. The global reconnection rate is substantially larger
as many independent patches come together. From Lazarian et al. 2004.}
\label{fig_rec}
 \end{center}
\end{figure}

LV99 we introduced a model that included the effects of magnetic field line wandering (see Figure 1). The model relies on the nature of three-dimentional magnetic field wandering in turbulence. This nature is different in three and two dimensions, which provides the major difference between the LV99 model and the earlier attempts to solve the problem of magnetic reconnection appealing to turbulence (Matthaeus \& Lamkin 1985). The effects of compressibility and heating which were thought to be important in the earlier studies (Matthaeus \& Lamkin 1985, 1986) do not play the role for the LV99 model either. The model is applicable to any weakly perturbed magnetized fluid, irrespectively, of the degree of plasma being collisional or collisionless (cf. Shay et al. 1998). 

Two effects are the most important for understanding of the nature of reconnection in LV99. First of all, in three dimensions bundles of magnetic field lines can enter the reconnection region and reconnection there independently (see Figure~1), which is in contrast to two dimensional picture where in Sweet-Parker reconnection the process is artificially constrained. Then, the nature of magnetic field stochasticity and therefore magnetic field wandering (which determines the outflow thickness, as illustrated in Figure~1) is very different in 2D and the real 3D world (LV99). In other words, removing artificial constraints on the dimensionality of the reconnection region and the magnetic field being absolutely straight, LV99 explores the real-world astrophysical reconnection.      

Our calculations in LV99 showed that the resulting reconnection rate is limited only by the width of the outflow region.  This proposal, called ``stochastic reconnection'', leads to reconnection speeds close to the turbulent velocity in the fluid.  More precisely, assuming isotropically driven turbulence characterized by an injection scale, $l$, smaller than the current sheet length, we find
\begin{equation}
V_{rec}\approx \frac{u_l^2}{V_A}\left(l/L\right)^{1/2}\approx u_{turb}\left(l/L\right)^{1/2}, 
\label{recon1}
\end{equation}
where $u_l$  is the velocity at the driving scale and $u_{turb}$ is the velocity of the largest eddies of the strong turbulent cascade. Note, that here "strong" means only that the eddies decay through nonlinear interactions in an eddy turn over time (see more discussion of the  LV99).  All the motions are weak in the sense that the magnetic field lines are only weakly perturbed.  

It is useful to rewrite this in terms of the power injection rate $P$. As the perturbations on the injection scale of turbulence are assumed to have velocities $u_l<V_A$, the turbulence is weak at large scales. Therefore, the relation between the power and the injection velocities are different from the usual Kolmogorov estimate, namely, in the case of the weak turbulence $P\sim u_l^4/(lV_A)$ (LV99). Thus we get,
\begin{equation}
V_{rec}\approx \left(\frac{P}{LV_A}\right)^{1/2} l,
\label{recon2}
\end{equation} 
where $l$ is the length of the turbulent eddies parallel to the large scale magnetic field lines as well as the injection scale.

The reconnection velocity given by equation (\ref{recon2}) does not depend on resistivity or plasma effects. Therefore for sufficiently high level of turbulence we expect both collisionless and collisional fluids to reconnect at the same rate.

\section{Testing of Lazarian \& Vishniac 99 Model}

%
\begin{figure*}
\center
\includegraphics[width=0.3\textwidth]{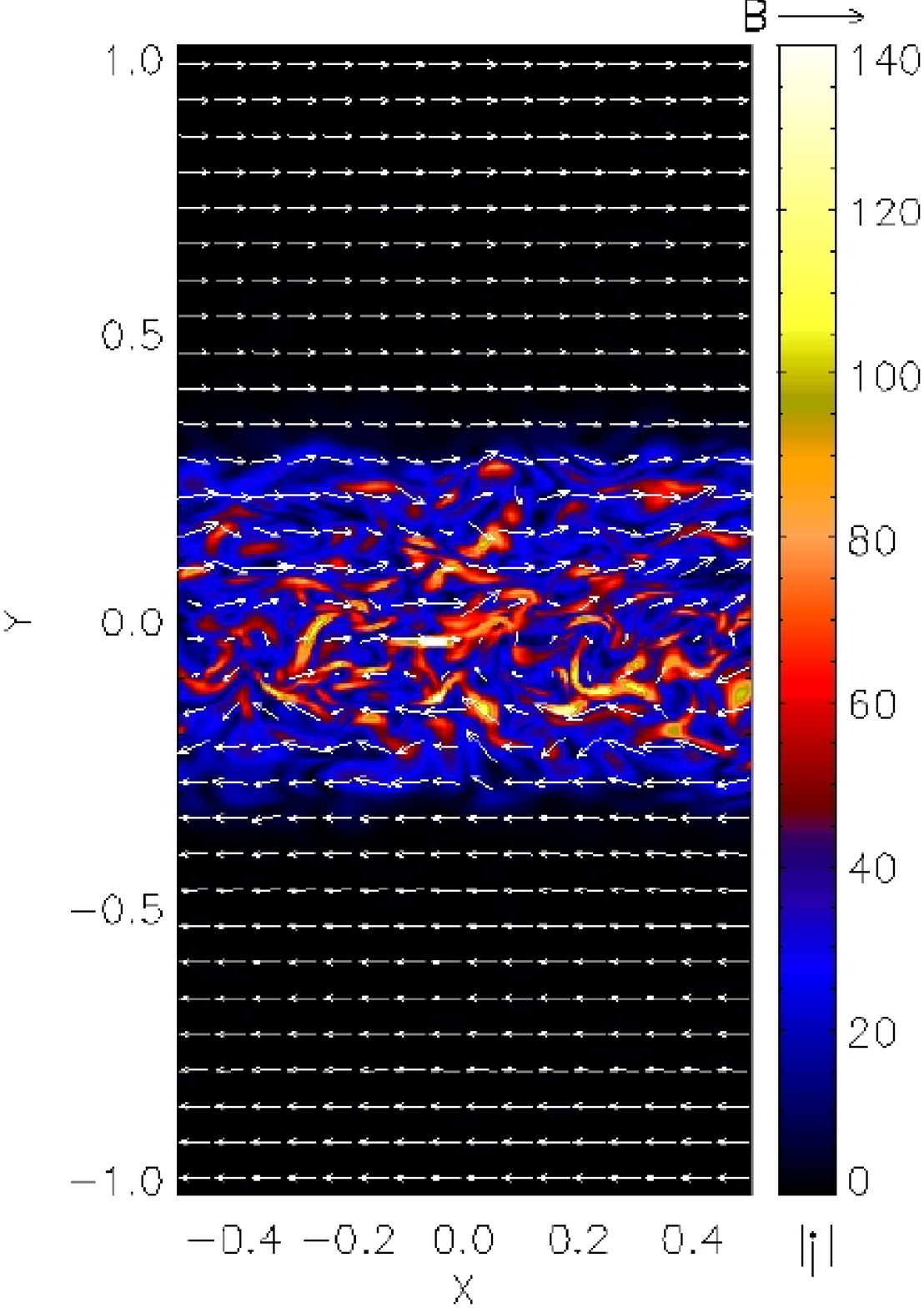}
\includegraphics[width=0.3\textwidth]{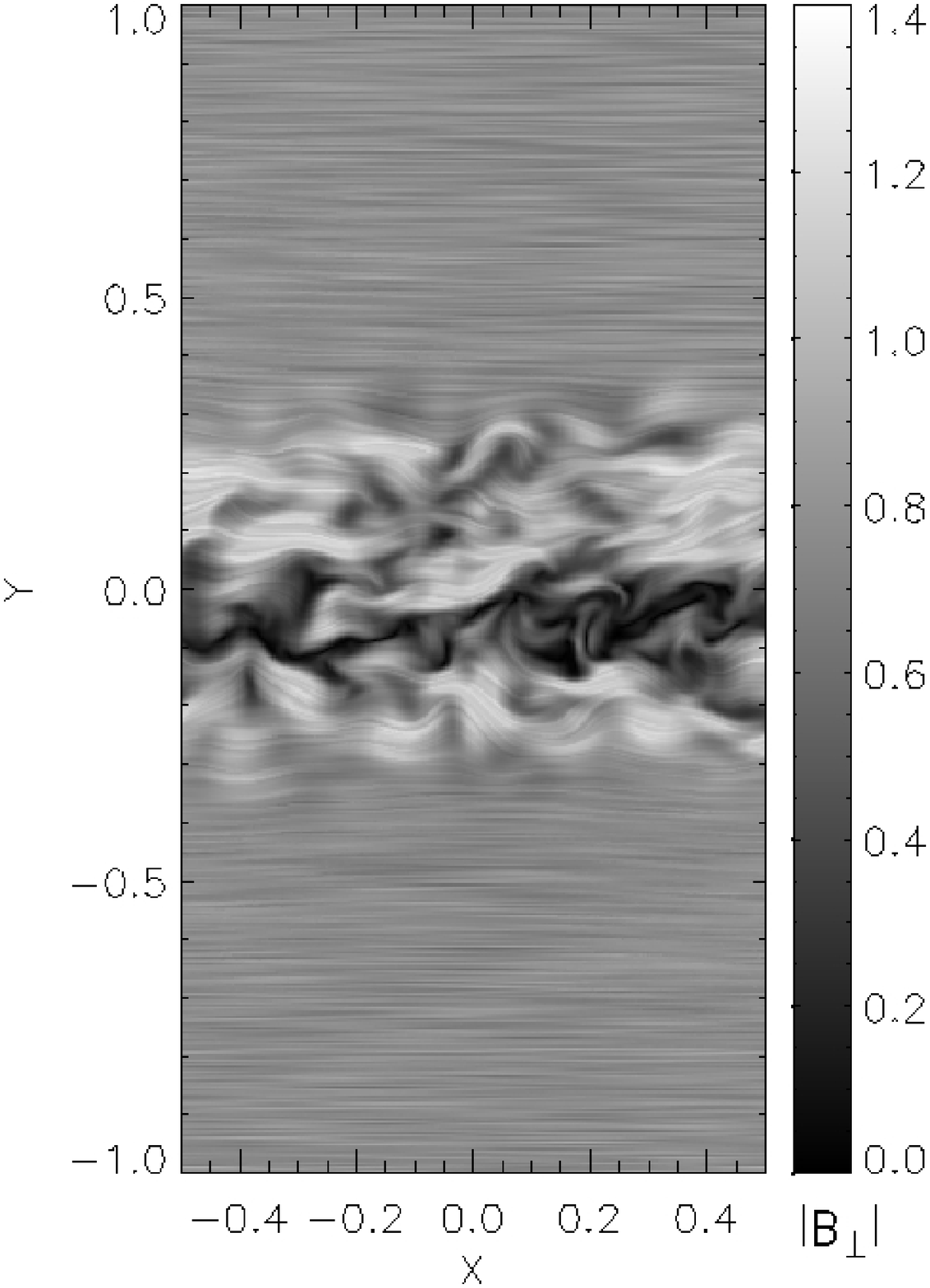}
\caption{{\it Left panel}: Current intensity and magnetic field configuration during stochastic reconnection.  We show a slice through the middle of the computational box in the xy plane after twelve dynamical times for a typical run.  The shared component of the field is perpendicular to the page.  The intensity and direction of the magnetic field is represented by the length and direction of the arrows.  The color bar gives the intensity of the current.  The reversal in $B_x$  is confined to the vicinity of y=0 but the current sheet is strongly disordered with features that extend far from the zone of reversal. {\it Right panel}: Representation of the magnetic field in the reconnection zone with textures.
\label{fig:top_turb}}
\end{figure*}

Here we describe the results of a series of three dimensional numerical simulations aimed at adding turbulence to the simplest reconnection scenario and testing equation (\ref{recon2}).  We take two regions with strongly differing magnetic fields lying next to one another.  The simulations are periodic in the direction of the shared field (the z axis) and are open in the reversed direction (the x axis).  The external gas pressure is uniform and the magnetic fields at the top and bottom of the box are taken to be the specified external fields plus small perturbations to allow for outgoing waves.  The grid size in the simulations varied from 256x512x256 to 512x1028x512 so that the top and bottom of the box are far away from the current sheet and the region of driven turbulence around it.   At the sides of the box where outflow is expected the derivatives of the dynamical variables are set to zero.  A complete description of the numerical methodology can be found in Kowal et al. (2009). All our simulations are allowed to evolve for seven Alfven crossing times without turbulent forcing.  During this time they develop the expected Sweet-Parker current sheet configuration with slow reconnection.  Subsequently we turn on isotropic turbulent forcing inside a volume centered in the midplane (in the xz plane) of the simulation box and extending outwards by a quarter of the box size.  The turbulence reaches its full amplitude around eight crossing times and is stationary thereafter.

In Figure 2 we see the current density on an xy slice of the computational box once the turbulence is well developed.  As expected, we see that the narrow stationary current sheet characteristic of Sweet-Parker reconnection is replaced by a chaotic structure, with numerous narrow peaks in the current density.  Clearly the presence of turbulence has a dramatic impact on the structure of the reconnection zone.  In addition, we see numerous faint features indicating weak reconnection between adjacent turbulent eddies. 

\begin{figure}
\center
\includegraphics[width=0.9\columnwidth]{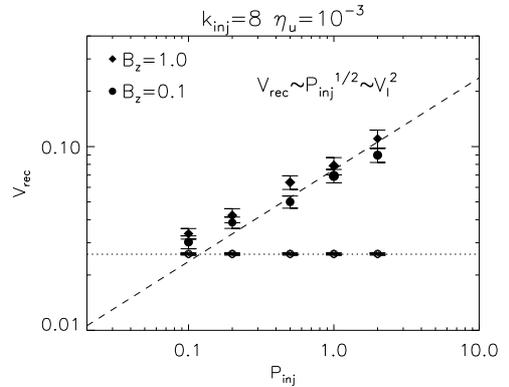}
\caption{Reconnection speed versus input power for the driven turbulence.  We show the reconnection speed, defined by equation (4) plotted against the input power for an injection wavenumber equal to 8 (i.e. a wavelength equal to one eighth of the box size) and a resistivity $\nu_u$.  The dashed line is a fit to the predicted dependence of  $P^{1/2}$ (see eq. (3)).  The horizontal line shows the laminar reconnection rates for each of the simulations before the turbulent forcing started.  Here the   uncertainty in the time averages are indicated by the size of the symbols and the variances are shown by the error bars.
\label{pow_dep}}
\end{figure}

The speed of reconnection in three dimensions can be hard to define without explicit evaluation of the magnetic field topology.  However, in this simple case we can define it as the rate at which the $x$ component of the magnetic field disappears.  More precisely, we consider a yz slice of the simulation, passing through the center of the box.  The rate of change of the area integral of  |$B_x$| is its flux across the boundaries of the box minus the rate at which flux is annihilated through reconnection (see more discussion in Kowal et al. 2009) 
\begin{equation}
\partial_t\left(\int|B_x|dzdy\right)=\oint sign(B_x)vec{E}d\vec{l}-2V_{rec}B_{x,ext}L_z
\label{measure}
\end{equation}
where electric field is $\vec{E}=\vec{v}\times \vec{B} -\eta \vec{j}$, $B_{x,ext}$ is the absolute value of $B_x$  far from the current sheet and $L_z$ is the width of the box in the $z$ direction.  This follows from the induction equation under the assumption that the turbulence is weak to lead to local field reversals and that the stresses at the boundaries are weak to produce significant field bending there.  In other words, fields in the $x$ direction are advected through the top and bottom of the box, and disappear only through reconnection.  Since we have assumed periodic boundary conditions in the $z$ direction the boundary integral on the right hand side is only taken over the top and bottom of the box.  By design this definition includes contributions to the reconnection speed from contracting loops, where Ohmic reconnection has occurred elsewhere in the box and $|B_x|$ decreases as the end of a reconnected loop is pulled through the plane of integration.  It is worth noting that this estimate is roughly consistent with simply measuring the average influx of magnetic field lines through the top and bottom of the computational box and equating the mean inflow velocity with the reconnection speed. Following equation (\ref{measure}) we can evaluate the reconnection speed for varying strengths and scales of turbulence and varying resistivity.

In Figure~\ref{pow_dep} we see the results for varying amounts of input power, for fixed resistivity and injection scale as well as for the case of no turbulence at all.  The line drawn through the simulation points is for the predicted scaling with the square root of the input power. The agreement between equation (\ref{recon2}) and Figure~\ref{pow_dep} is encouraging but does not address the most important aspect of stochastic reconnection, i.e. its insensitivity to $\eta$.

%
\begin{figure}
\center
\includegraphics[width=0.9\columnwidth]{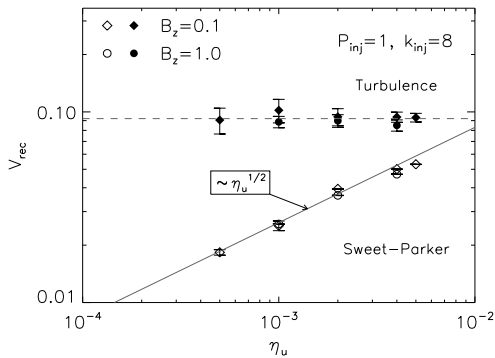}
\caption{Reconnection speed versus resistivity.  We show the reconnection speed plotted against the uniform resistivity of the simulation for an injection wavenumber of 8 and an injected power of one.  We include both the laminar reconnection speeds, using the hollow symbols, fit to the expected dependence of $\eta_u$, and the stochastic reconnection speeds, using the filled symbols.  As before the symbol sizes indicate the uncertainty in the average reconnection speeds and the error bars indicate the variance.  We included simulations with large, $B_z=1$, and small, $B_z=0.1$, guide fields.
\label{ueta_dep}}
\end{figure}

In Figure~\ref{ueta_dep} we plot the results for fixed input power and scale, while varying the background resistivity.  In this case $\eta$  is taken to be uniform, except near the edges of the computational grid where it falls to zero over five grid points.  This was done to eliminate edge effects for large values of the resistivity. We see from the Figure~\ref{ueta_dep} that the points for laminar reconnection scale as $\sqrt{\eta}$, the expected scaling for Sweet-Parker reconnection.  In contrast, the points for reconnection in a turbulent medium do not depend on the resistivity at all. In summary, we have tested the model of stochastic reconnection in a simple geometry meant to approximate the circumstances of generic magnetic reconnection in the universe.  Our results are consistent with the mechanism described by LV99.  The implication is that turbulent fluids in the universe including the interstellar medium, the convection zones of stars, and accretion disks, have reconnection speeds close to the local turbulent velocity, regardless of the local value of resistivity.  Magnetic fields in turbulent fluids can change their topology on a dynamical time scale. 

In Kowal et al. (2009) we also studied the dependence of the reconnection on the anomalous resistivity, which increases effective resistivity for high current densities. The anomalous resistivity can be used as a proxy for plasma effects, e.g. collisionless effects in reconnection. While it enhances the local speed of individual reconnection events results in Kowal et al. (2009) testify that the total reconnection rate does not change.

\begin{figure}
\center
\includegraphics[width=0.9\columnwidth]{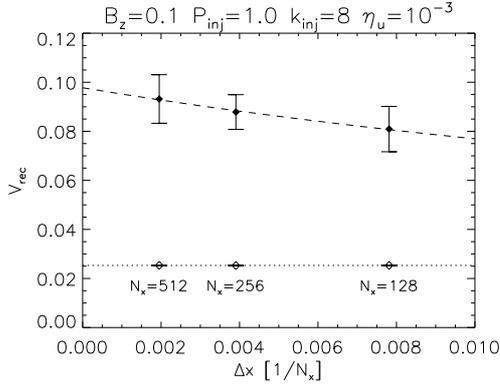}
\caption{Dependence of the reconnection rate on the numerical resolution. If the fast reconnection were due to yet unclear numerical effects on small scales, we would expect to see the increase of the reconnection rate with the decrease of the numerical box. If anything, the actual dependence of the reconnection rate on the box size shows the opposite dependence.
\label{fig:reso_dep}}
\end{figure}

Any numerical study has to address the issue of the possible numerical effects on the results. We show the dependence of the reconnection rate on the numerical resolution in Figure~\ref{fig:reso_dep}. The reconnection rate increases with the increase of the resolution, which testifies that the fast reconnection is not due to numerical effects. Indeed, higher numerical reconnection is expected for lower resolution simulations.

%

Finally, it is important to give a few words in relation to our turbulence driving. We drive our turbulence solenoidally to minimize the effects of compression, which does not play a role in LV99 model. The turbulence driven in the volume around the reconnection layer corresponds to the case of astrophysical turbulence, which is also volume-driven. On the contrary, the case of the turbulence driven at the box boundaries would produce spatially inhomogeneous imbalanced turbulence
for which we do not have analytical predictions (see discussion of such turbulence in Beresnyak \& Lazarian 2009). We stress, that it is not the shear size of our numerical simulations, but the correspondence of the observed scalings to those predicted in LV99 that allows us to claim that we proved that the 3D reconnection is fast in the presence of turbulence. 

\section{Acceleration of Cosmic Rays}

\begin{figure}[!t]
\includegraphics[width=\columnwidth]{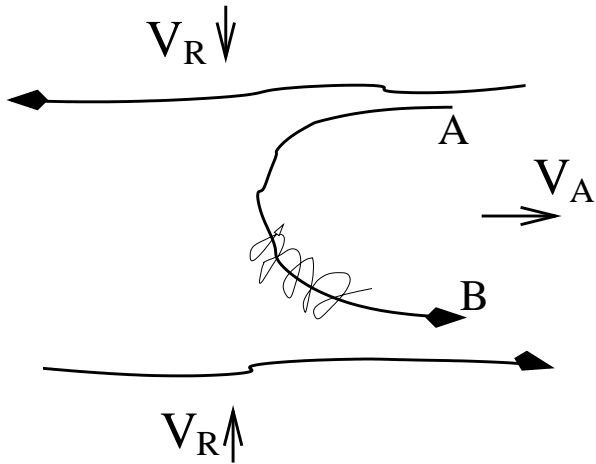}
\caption{  
Cosmic rays spiral about a reconnected magnetic
field line and bounce back at points A and B. The reconnected
regions move towards each other with the reconnection velocity
$V_R$. The advection of cosmic rays entrained on magnetic field
lines happens at the outflow velocity, which is in most cases
of the order of $V_A$. Bouncing at points A and B happens
because either of streaming instability induced by energetic particles or magnetic
turbulence in the
reconnection region. In reality, the outflow region gets filled in by the oppositely moving tubes of reconnected flux which collide only to repeat on a smaller scale the pattern of the larger scale reconnection.  From Lazarian (2005).}
\label{fig_recon}
\end{figure} 

In what follows we discuss the first order Fermi acceleration which arises from volume-filling reconnection\footnote{We would like to stress that Figure 1 exemplifies only the first moment of reconnection when the fluxes are just brought together. As the reconnection develops the volume of thickness $\Delta$ gets filled with the reconnected 3D flux ropes moving in the opposite directions.}. The LV99 presented such a model of reconnection and observations of the Solar magnetic field reconnection support the volume-filled idea (Ciaravella \& Raymond 2008).

Figure~\ref{fig_recon} exemplifies the simplest realization of the acceleration within the reconnection region expected within LV99 model. As a particle bounces back and forth between converging magnetic fluxes, it gains energy through the first order Fermi acceleration described in de Gouveia dal Pino \& Lazarian (2003, 2005, henceforth GL05) (see also Lazarian 2005).

To derive the energy spectrum of particles one can use the routine way of dealing with the first order Fermi acceleration in shocks (see Longair 1992). Consider the process of acceleration of $M_0$ particles with the initial energy $E_0$. If a particle gets energy $\beta E_0$ after a collision, its energy after $m$ collisions is $\beta^m E_0$. At the same time if the probability of a particle to remain within the accelerating region is $P$, after $m$ collisions the number of particles gets $P^m M_0$. Thus $\ln (M/M_0)/\ln(E/E_0)=\ln P/\ln\beta$ and
\begin{equation}
\frac{M}{M_0}=\left(\frac{E}{E_0}\right)^{\ln P/\ln\beta}
\end{equation}
For the stationary state of accelerated particles the number $M$ is the number of particles having energy equal or larger than $E$, as some of these particles are not lost and are accelerated further. Therefore:
\begin{equation}
N(E)dE=const\times E^{-1+(\ln P/\ln\beta)} dE
\label{NE}
\end{equation}

To determine $P$ and $\beta$ consider the following process. The particles from the upper reconnection region see the lower reconnection region moving toward them with the velocity $2V_{R}$ (see Figure~\ref{recon2}). If a particle from the upper region enters at an angle $\theta$ into the lower region the expected energy gain of the particle is $\delta E/E=2V_{R}\cos\theta/c$. For isotropic distribution of particles their probability function is $p(\theta)=2\sin\theta\cos\theta d\theta$ and therefore the average energy gain per crossing of the reconnection region is
\begin{equation}
\langle \delta E/E \rangle =\frac{V_{R}}{c}\int^{\pi/2}_{0} 2\cos^2\theta \sin\theta d\theta=4/3\frac{V_{R}}{c}
\end{equation}
An acceleration cycle is when the particles return back to the upper reconnection region. Being in the lower reconnection region the particles see the upper reconnection region moving with the speed $V_{R}$. As a result, the reconnection cycle provides the energy increase $\langle \delta E/E \rangle_{cycle}=8/3(V_{R}/c)$ and
\begin{equation}
\beta=E/E_0=1+8/3(V_{R}/c)
\label{beta}
\end{equation}

Consider the case of $V_{diff}\ll V_R$. The total number of particles crossing the boundaries of the upper and lower fluxes is $2\times 1/4 (n c)$, where $n$ is the number density of particles. With our assumption that the particles are advected out of the reconnection region with the magnetized plasma outflow the loss of the energetic particles is $2\times V_{R}n$. Therefore the fraction of energetic particles lost in a cycle is $V_{R} n/[1/4(nc)]=4V_{R}/c$ and
\begin{equation}
P=1-4V_{R}/c.
\label{P}
\end{equation}

Combining Eq.~(\ref{NE}), (\ref{beta}), (\ref{P}) one gets
\begin{equation}
N(E)dE=const_1 E^{-5/2}dE,
\label{-5/2}
\end{equation}
which is the spectrum of accelerated energetic particles for the case when the back-reaction is negligible (see also GL05)\footnote{The obtained spectral index is similar to the one of Galactic cosmic rays.}.

The first order acceleration of particles entrained on the contracting magnetic loop can be understood from the Liouville theorem. As in the process of the magnetic tubes are contracting, the regular increase of the particle's energies is expected. The requirement for the process to proceed efficiently is to keep the accelerated particles within the contracting magnetic loop. This introduces limitations on the particle diffusivities perpendicular to magnetic field direction.   
The subtlety of the point above is related to the fact that while in the first order Fermi acceleration in shocks magnetic compression is important, the acceleration via LV99 reconnection process is applicable to incompressible fluids. Thus, unlike shocks, not the entire volume that shrinks for the acceleration, but only the volume of the magnetic flux tube. Thus high perpendicular diffusion of particles may decouple them from the magnetic field. Indeed, it is easy to see that while the particles within a magnetic flux rope depicted in Figure~6 bounce back and forth between the converging mirrors and get accelerated, if these particles leave the flux rope fast, they may start bouncing between the magnetic fields of different flux ropes which may sometimes decrease their energy. Thus it is important that the particle diffusion parallel and perpendicular magnetic field stays different. Particle anisotropy which arises from particle preferentially getting acceleration in terms of the parallel momentum may also be important.

\section{Simulations of the Acceleration of Cosmic Rays by Reconnection}

In the numerical studies of the cosmic ray acceleration we use data cubes obtained from the models of the weakly stochastic magnetic reconnection described in \S 2.  For a given snapshot we obtain a full configuration of the plasma flow variables (density and velocity) and magnetic field.  We inject test particles in such an environment and integrate their trajectories solving the motion equation for relativistic charged particles
\begin{equation}
 \frac{d}{d t} \left( \gamma m \vec{u} \right) = q \left( \vec{E} + \vec{u} \times \vec{B} \right) ,
\end{equation}
where $\vec{u}$ is the particle velocity, $\gamma \equiv \left( 1 - u^2 / c^2 \right)^{-1}$ is the Lorentz factor, $m$ and $q$ are particle mass and electric charge, respectively, and $c$ is the speed of light.

The study of the magnetic reconnection is done using the magnetohydrodynamic fluid approximation, thus we do not specify the electric field $\vec{E}$ explicitly.  Nevertheless, the electric field is generated by either the flow of magnetized plasma or the resistivity effect and can be obtained from the Ohm's equation
\begin{equation}
 \vec{E} = - \vec{v} \times \vec{B} + \eta \vec{j} ,
\end{equation}
where $\vec{v}$ is the plasma velocity and $\vec{j} \equiv \nabla \times \vec{B}$ is the current density.

In our studies we are not interested in the acceleration by the electric field resulting from the resistivity effects, thus we neglect the last term.  After incorporating the Ohm's law, the motion equation can be rewritten as
\begin{equation}
 \frac{d}{d t} \left( \gamma m \vec{u} \right) = q \left[ \left( \vec{u} - \vec{v} \right) \times \vec{B} \right] . \label{eq:trajectory}
\end{equation}

In our simulation we do not include the particle energy losses, so particle can gain or loose through the interaction with moving magnetized plasma only.  For the sake of simplicity, we assume the speed of light 20 times larger than the Alfven speed $V_A$, which defines plasma in the nonrelativistic regime, and the mean density is assumed to be 1 atomic mass unit per cubic centimeter, which is motivated by the interstellar medium density.  We integrate equation~\ref{eq:trajectory} for 10,000 particles with randomly chosen initial positions in the domain and direction of the motion.

\begin{figure}[ht]
 \center
 \includegraphics[width=0.8\columnwidth]{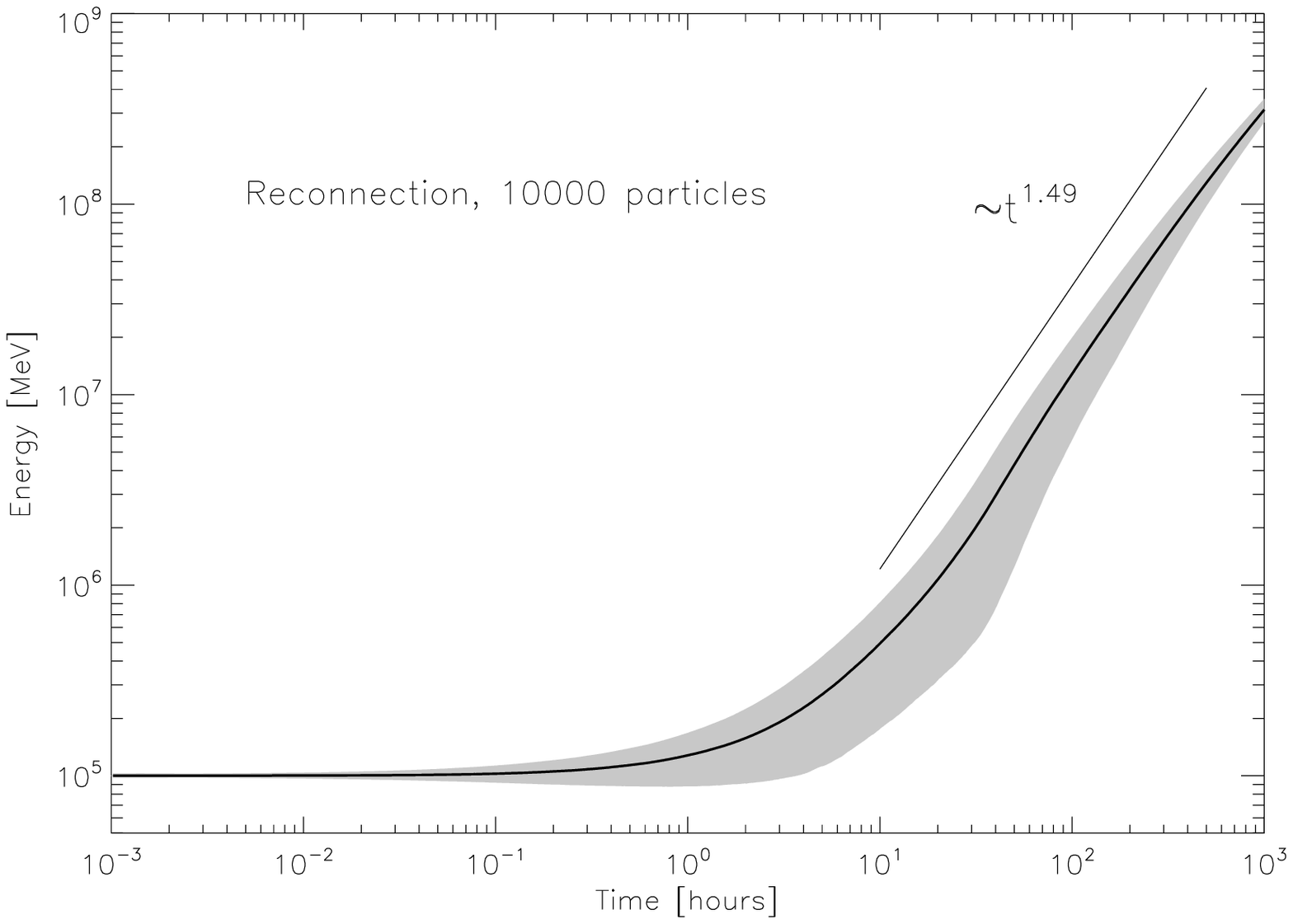}
 \includegraphics[width=0.8\columnwidth]{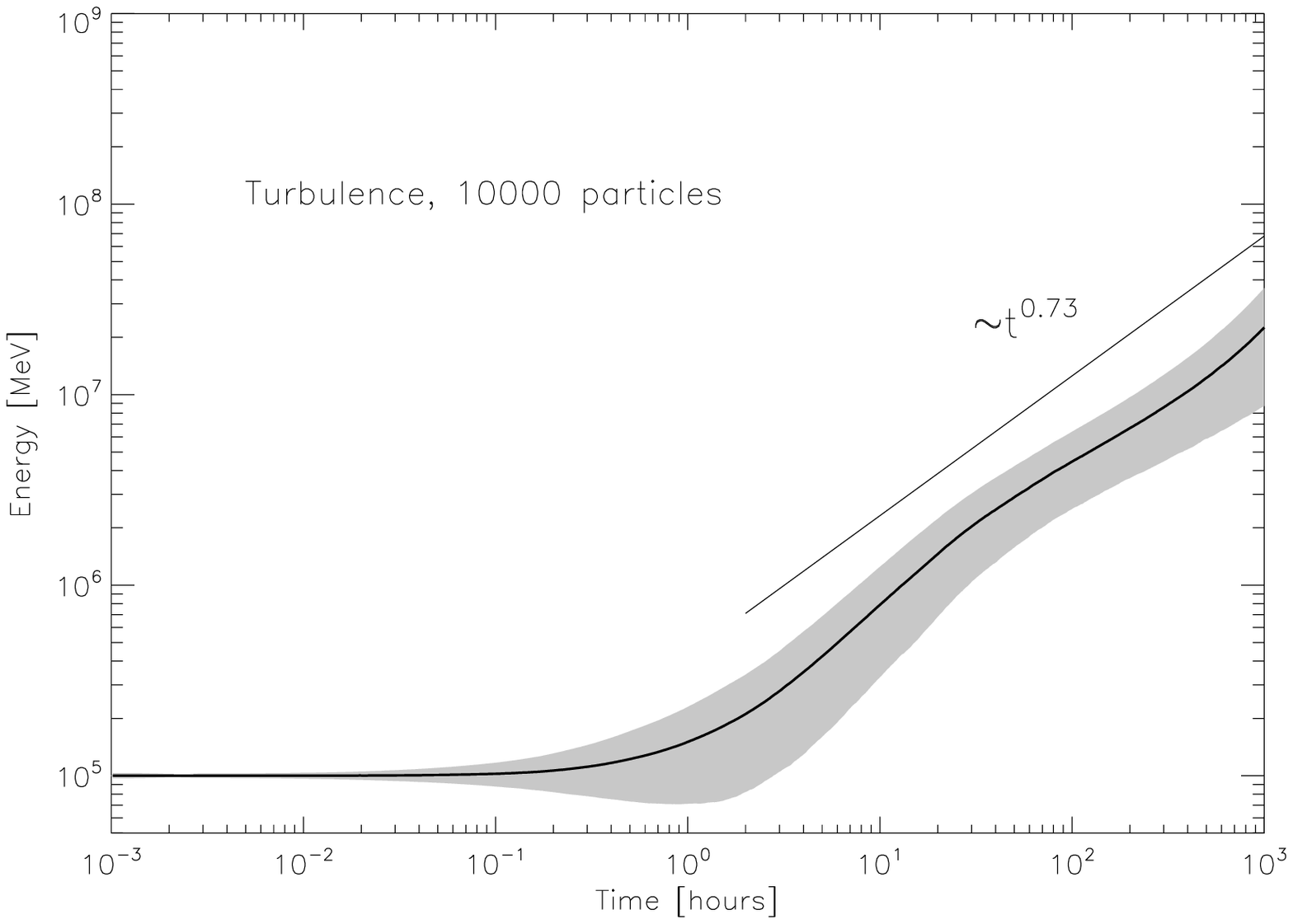}
 \caption{Particle energy evolution averaged over 10,000 particles with the initial energy $E_0=10^5$ MeV and random initial positions and directions.  In the upper plot we show results for the weakly stochastic turbulence environment and in the lower plot for turbulent environment without magnetic reconnection.  In both cases we assume $c = 20 V_A$ and $\langle \rho \rangle = 1$ u/cm$^{3}$. \label{fig:energy}}
\end{figure}

In Figure~\ref{fig:energy} we show the particle energy evolution averaged over all integrated particles for two cases. In the left plot we used plasma fields topology obtained from the weakly stochastic magnetic reconnection models, in the right plot we use fields topology taken from the turbulence studies.  Gray area shows the particle energy dispersion over the group of particles.

In the case of reconnection model, the expected exponential acceleration is observed until time about 100 hours.  Later on, the physical limitations of the computational domain result in a different growth rate corresponding to $E\sim t^{1.49}$.  In the case of turbulence without large scale magnetic reconnection the growth of energy is slower $E\sim t^{0.73}$.  This testifies that that the presence of reconnection makes the acceleration more efficient. The numerical confirmation of the first order acceleration in the reconnection regions is presented in our forthcoming paper.

\begin{figure}[ht]
 \center
 \includegraphics[width=0.8\columnwidth]{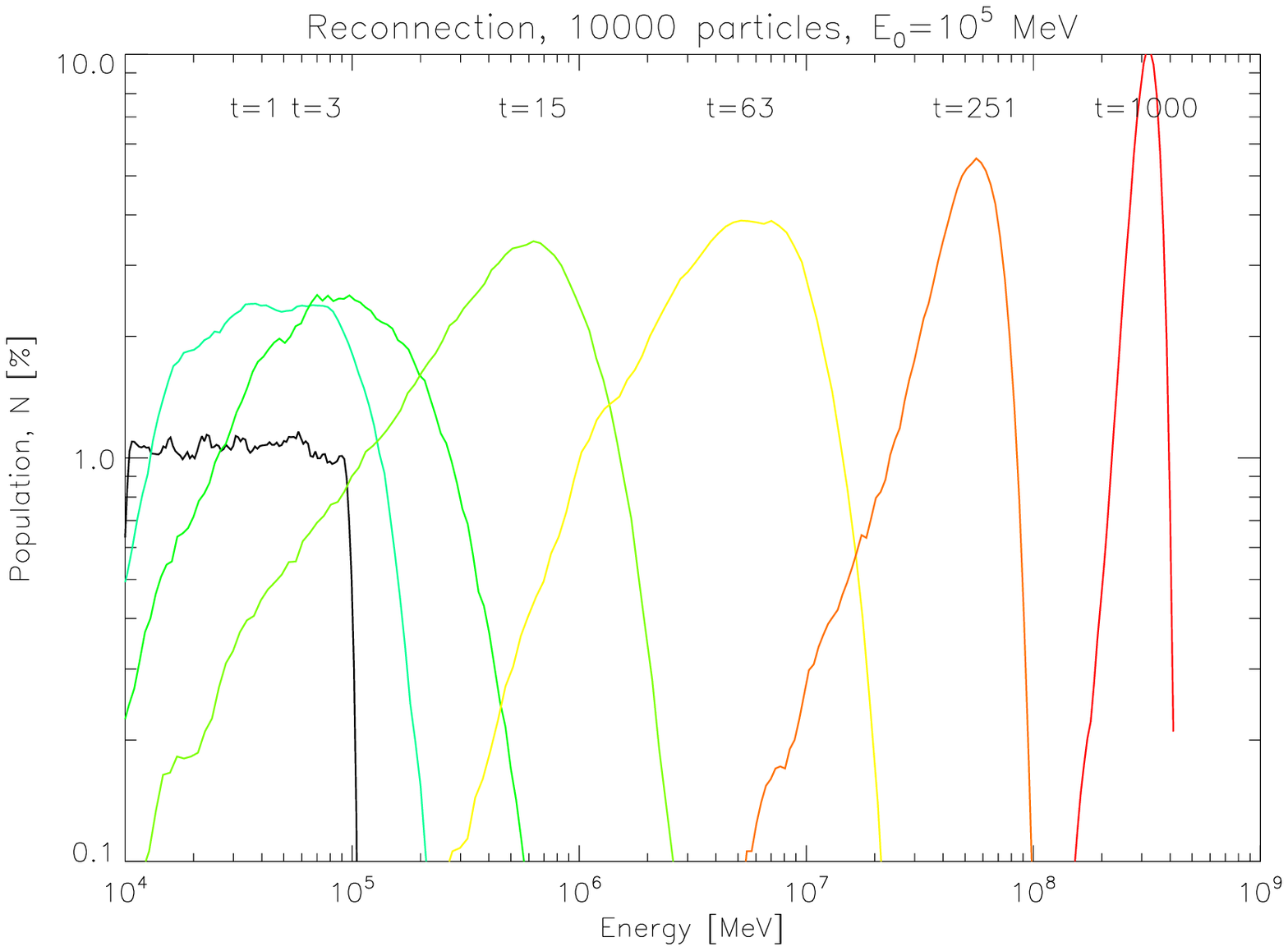}
 \includegraphics[width=0.8\columnwidth]{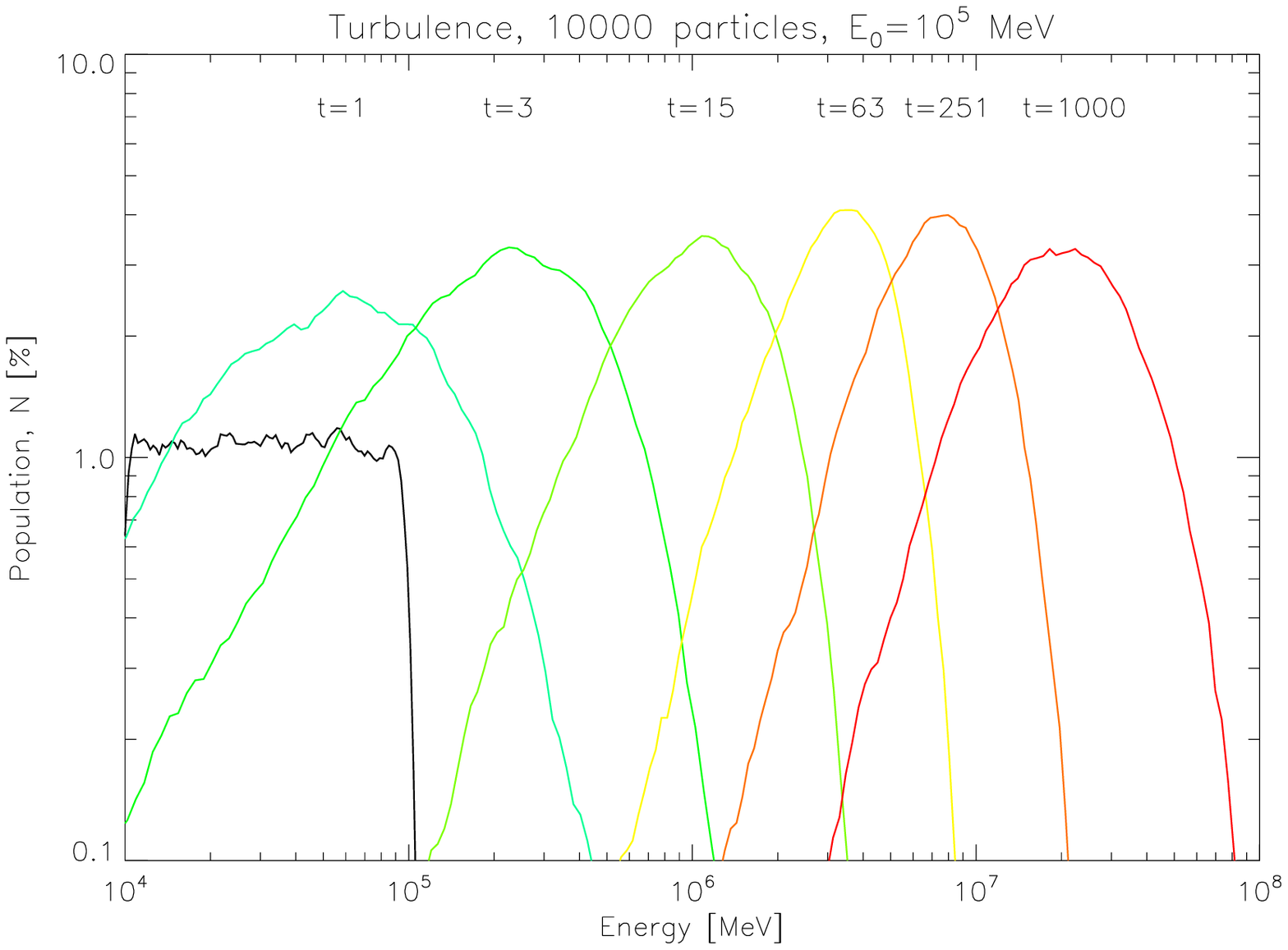}
 \caption{Particle spectrum evolution for 10,000 particles with the uniform initial energy distribution $E_0=10^5-10^6$ MeV and random initial positions and directions.  In the upper plot with show results for the weakly stochastic turbulence environment and in the lower plot for turbulent environment without magnetic reconnection.  In both cases we assume $c = 20 V_A$ and $\langle \rho \rangle = 1$ u/cm$^{3}$. \label{fig:spectrum}}
\end{figure}
In Figure~\ref{fig:spectrum} we show the evolution of particle energy distribution.  Initially uniform distribution of particle energy ranging from $10^5$ to $10^6$~MeV evolves faster to higher energies if the reconnection is present.  In this case, the final distribution is log-normal being more peaked over the time and with decreasing dispersion of energies in logarithmic scale.  On the contrary, in the case of pure turbulence, the energy distribution after evolving to the log-normal shape preserves its dispersion over the time. 

As magnetic reconnection is ubiquitous process, the particle acceleration within reconnection process should be widely spread. Therefore accepting the preliminary character of these results above
we are involved in more extensive studies of the acceleration-via-reconnection.

\section{Explanation of the Anomalous Cosmic Ray Origin}

The processes of the energetic particle acceleration in the process of turbulent reconnection can preaccelerate particles to the intermediate energies helping to solve the problem of particle injection into shocks. It can also act as the principal process of acceleration. Below we present the case where we believe that the latter takes place. 

Since the crossing of the termination shock (TS) by Voyager 1 (V1) in late 2004 and by Voyager 2 (V2) in mid 2007 it became clear that several paradigms needed to be revised. Among them was the acceleration of particles. Prior to the encounter of the termination shock by V1 the prevailing view was that anomalous cosmic rays (ACRs) were accelerated at the TS by diffusive shock acceleration (DSA) to energies 1-300 MeV/nuc (e.g., Jokipii \& Giacalone, 1998; Cummings \& Stone, 1998). However, with the crossing of the TS by V1 the energy spectrum of ACR did not unroll to the expected source shape: a power-law at lower energies with a roll off at higher energies. After 2004, both the V1 spectrum in the heliosheath and the V2 spectrum upstream the TS, continued to evolve toward the expected source shape.

To explain this paradox several models were proposed. Among them, McComas \& Schwadron (2006) suggested that at a blunt shock the acceleration site for higher energy ACRs would be at the flanks of the TS, where the injection efficiency would be higher for DSA and connection times of the magnetic field lines to the shock would be longer, allowing acceleration to higher energies. Fisk et al. (2006) on the other hand suggested that stochastic acceleration in the turbulent heliosheath would continue to accelerate ACRs and that the high-energy source region would thus be beyond the TS. Other works, such as Jokipii (2006) and Florinki and Zank (2006) try to explain the deficit of ACRs based on a dynamic termination shock. Jokipii (2006) pointed out that a shock in motion on time scales of the acceleration time of the ACRs, days to months, would cause the spectrum to differ from the expected DSA shape. Florinki \& Zank (2006) calculated the effect of Magnetic Interacting Regions (MIRs) with the Termination Shock on the ACR spectral shape. They show that there is a prolonged period of depressed intensity in mid-energies from a single MIR. Other recent works have included stochastic acceleration, as well as other effects (Moraal et al., 2006, 2007; Zhang, 2006; Langner and Potgieter, 2006; Ferreira et al., 2007). It became clear after the crossing of the TS by V2 that these models would require adjustments. The observations by V2 indicate for example that a transient did not cause the modulation shape of the V2 spectrum at the time of its TS crossing. When both spacecraft were in the heliosheath in late 2007, the radial gradient in the 13-19 MeV/nuc ions did not appear to be caused by a transient. The 60-74 MeV/nuc ions have no gradient, so no north-south or longitudinal asymmetry is observed in the ACR intensities at the higher energies.

In Lazarian \& Opher (2009, LO09) we propose an alternative model, which explains the source of ACRs as being in the heliosheath and we appeal to magnetic reconnection as a process that can accelerate particles. LO09  explained the origin of the magnetic field reversals that induce magnetic reconnection in heliosheath and heliopause.

Indeed, it is well known that magnetic field in the heliosphere change
polarity and induce reconnection. For instance,
as the Sun rotates magnetic field twists into a Parker spiral (Parker
1958) with magnetic fields separated by a current sheet (see Schatten
1971). The changes of magnetic field are also expected due to the
Solar cycle activity.

\begin{figure}[ht]
 \center
 \includegraphics[width=0.7\columnwidth]{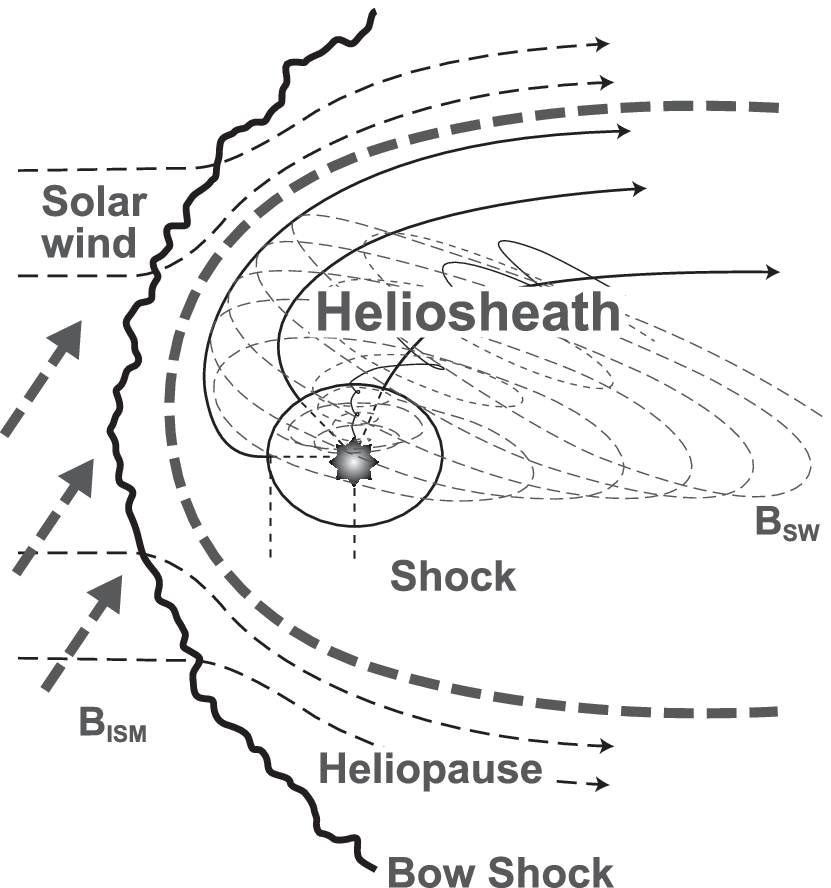}
 \includegraphics[width=0.7\columnwidth]{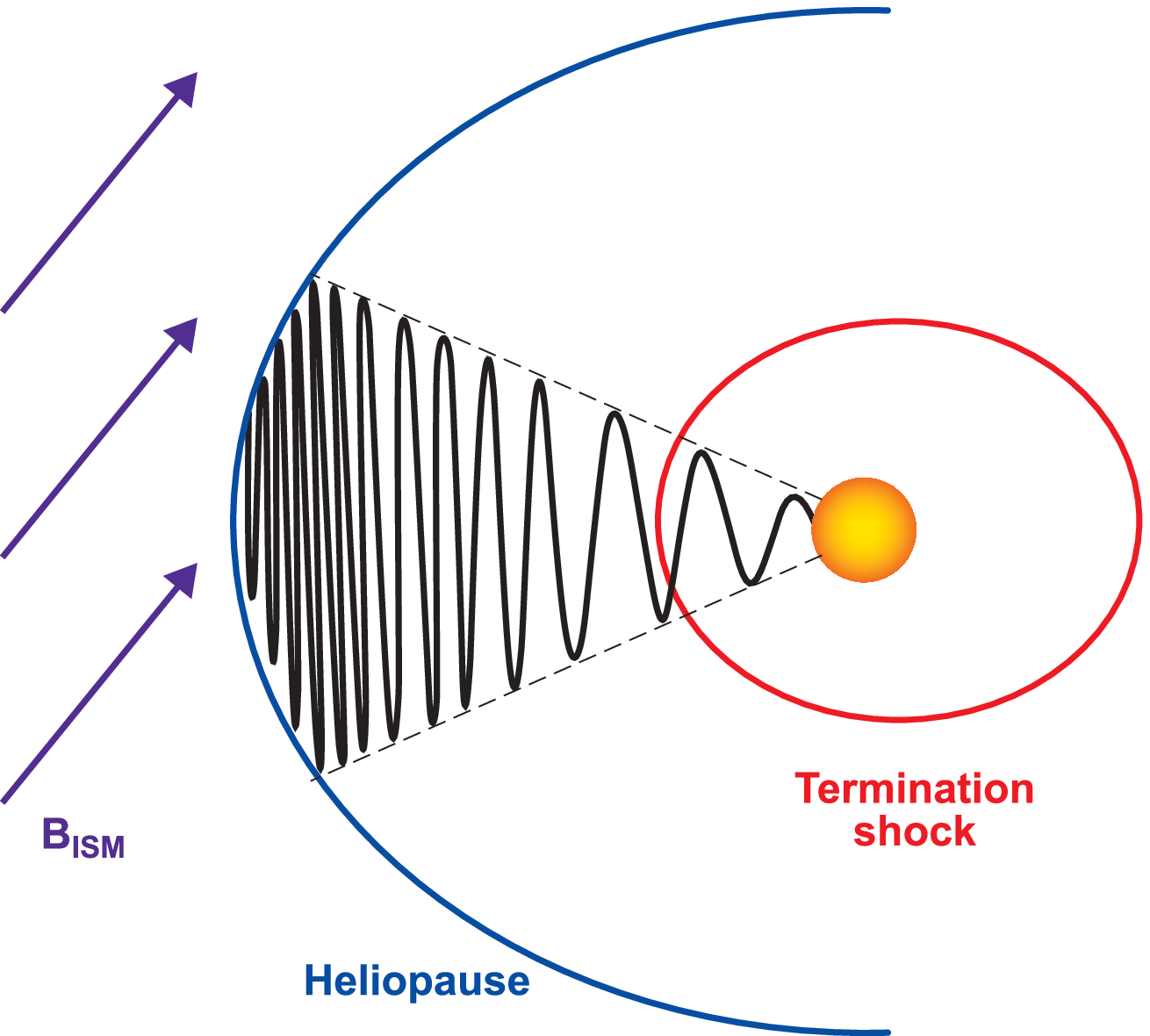}
 \caption{{\it Upper plot}. Global view of the interaction of the solar wind with the interstellar wind. The spiral solar magnetic field (shown in dark dashed lines) is  shown being deflected at the heliopause. The heliopause itself is being deflected by the interstellar magnetic field. (figure adapted from S. Suess (2006). {\it Lower plot}. A meridional view of the boundary sectors of the 
heliospheric currenty sheet  and how the opposite sectors get tighter 
closer to the heliopause. The thickness of the outflow regions in the 
reconnection region depends on the level of turbulence. From LO09. \label{fig_anomal}}
\end{figure}

The question now is at what part of the heliosheath we expect to see
reversals. The structure of the magnetic field in the solar wind is
complex. The solar magnetic field lines near the termination shock are
azimuthal and form a spiral (see Figure \ref{fig_anomal}). We expect the 
reconnection and the corresponding energetic particle acceleration to happen at 
the heliosheath closer to the heliopause. This explains why Voyagers do not see
the signatures of anomalous cosmic ray acceleration as they pass the termination shock. 
Appealing to their model of collisionless reconnection, Drake et al. (2010) provided a similar
explanation of the origin of the anomalous cosmic rays.

\section{Convergence with other Models of Reconnection and Acceleration}

Since the introduction of the LV99 model, more traditional approaches to reconnection have been changed considerably. At the time of its introduction, the models competing with LV99 were some modifications of a single X-point collisionless reconnection. Those models had point-wise localized reconnection region and inevitably prescribed opening of the reconnection region upon the scales comparable to $L$ (see Figure~1). Such reconnection was difficult to realize in astrophysical conditions in the presence of random forcing which at high probability would collapse the opening of the reconnection layer. Single X-point reconnection were rejected in observations of solar flares by Ciaravella \& Raymond (2008).

Modern models of collisionless reconnection resemble the original LV99 model in a number of respects. For instance, they discuss, similarly to the LV99, the volume filled reconnection, although one may still wonder how this volume filling is being achieved in the presence of a single reconnection layer (see Drake et al. 2006).  While the authors still talk about islands produced in the reconnection, in three dimensions these islands are expected to evolve into contracting 3D loops or ropes (Daughton et al. 2008), which is similar to what is depicted in Figure~\ref{fig_recon}. Thus we do not expect to see a cardinal difference between the first order Fermi processes of the acceleration described in GL03 and later in Drake et al. (2006). This suggests that the backreaction of the particles calculated in Drake et al. (2006) considering the firehose instability may be employed as a part of the acceleration process described in GL03. 

The departure from the idea of regular reconnection and introduction of magnetic stochasticity is also obvious in a number of the recent papers appealing to the tearing mode instability\footnote{The idea of appealing to the tearing mode as a means of enhancing the reconnection speed can be traced back to Strauss (1988), Waelbroeck (1989) and Shibata \& Tanuma (2001). LV99 showed that the linear growth of tearing modes is insufficient to obtain fast reconnection. The new attack on the problem assumes that the non-linear growth of the islands due to merging provides their growth rates at the large scales that are larger than the direct growth of the tearing modes at those scales. This situation when the non-linear growth is faster than the linear one is rather unusual and requires further investigation.} as the process of enhancing reconnection (Loureiro et al. 2009, Bhattacharjee et al. 2009). The 3D loops that should arise as a result of this process should be able to accelerate energetic particles via the process described in GL03. As tearing modes can happen in a collisional fluid, this may potentially open another channel of reconnection in such fluid. The limitation of this process is that the tearing mode reconnection should not be too fast as this would present problems with explaining the accumulation of the flux prior to the flare. At the same time the idea of tearing reconnection does not have the natural valve of enhancing the reconnection speed, which is contrary to the LV model where the degree of reconnection is determined by the level of turbulence. Thus the periods of slow reconnection in LV99 model are ensured by the low level of turbulence prior to the flare. We believe that tearing reconnection can act to distabilize the Sweet-Parker reconnection layer, inducing turbulence.

We note, however, that in most astrophysical situations one has to deal with the {\it pre-existing} turbulence, which is the consequence of high Reynolds number of the fluid. Such turbulence may modify or suppress instabilities, including the tearing mode instability. We claim that it, by itself, induces fast reconnection. We may also argue that even if the astrophysical fluid is kept initially laminar the fluid a thick outflow from the reconnection region caused by tearing is expected to become turbulent. We suspect that this may be the cause of the reconnection explosions reported recently (Lapenta 2008, Bettarini \& Lapenta 2009).    

\section{Summary}

The successful testing of the LV99 model of fast reconnection opens avenues for the search of implications of that scheme. One of the implications of the model is the first order Fermi acceleration of energetic particles in the reconnection layer. As reconnection processes are expected to be ubiquitous in astrophysics, we expect the acceleration in reconnection layers to be also ubiquitous. Our simulations of the energetic particle acceleration in the reconnection layer provide results consistent with the first order Fermi acceleration. The origin of the anomalous cosmic rays may be related with the mechanism of particle acceleration via reconnection.

{\bf Acknowledgments}.
A.L. acknowledges NSF grants AST 0808118 and ATM 0648699, as well as the support of the NSF Center for Magnetic Self-Organization.


\bibliographystyle{elsarticle-harv}
\bibliography{<your-bib-database>}



\end{document}